# Graphene nanopipette enabled liquid delivery at zeptoliter precision

Shi Qiu[1], Yu Chen[2], Gediminas Gervinskas[3], Ross K.W. Marceau[4], Changxi Zheng[5,6*], Gang Sha[7*], Jing Fu[1,8*]

[1]Department of Mechanical and Aerospace Engineering, Monash University, Clayton, VIC 3800, Australia

[2]Monash Centre for Electron Microscopy, Monash University, Clayton, VIC 3800, Australia

[3]Monash Ramaciotti Centre for Cryo Electron Microscopy, Monash University, Clayton, VIC 3800, Australia

[4]Deakin University, Institute for Frontier Materials, Geelong, VIC 3216, Australia

[5]School of Science, Westlake University, 18 Shilongshan Road, Hangzhou 310024, Zhejiang Province, China

[6]Institute of Natural Sciences, Westlake Institute for Advanced Study, 18 Shilongshan Road, Hangzhou 310024, Zhejiang Province, China.

[7]School of Materials Science and Engineering/Herbert Gleiter Institute of Nanoscience,

Nanjing University of Science and Technology, Xiaolingwei 200, Nanjing, Jiangsu, 210094 China

[8]ARC Centre of Excellence for Advanced Molecular Imaging, Monash University, Clayton, VIC 3800, Australia

*Corresponding authors. Email: zhengchangxi@westlake.edu.cn or gang.sha@njust.edu.cn or jing.fu@monash.edu

**Abstract**

Accurate extraction of liquid is the first step towards low-volume liquid delivery and nanocharacterization, which plays a significant role in biomedical research. In this study, a tip-shaped graphene nanopipette (GNP) is proposed by encapsulating the biomolecule solution on the prefabricated metal tip with graphene. The volume of the encapsulated liquid is highly controllable at zeptoliter precision by tuning the encapsulating speed and the number of graphene encapsulation rounds. Using protein (ferritin) solution as an example, it has been confirmed by finite element analysis and the controlled experiments that the GNP allows the delivery of ferritin solution at the zeptoliter-scale. Furthermore, GNP is demonstrated as a new type of tip-shaped liquid cell, which is suitable for multiple nanocharacterization techniques. In particular, due to the ultra-sharp tip shape, isotope ($^{13}$C)-labelled glucose solution encapsulated in GNP has been characterized by atom probe tomography (APT) in the laser-pulsed mode. Analysis of the mass spectrum and the reconstructed three-dimensional chemical maps reveals the quantitative distribution and the compositions of individual glucose molecules. The GNP is expected to be introduced to deliver liquid in the range of zeptoliters to attoliters, and brings a new capability for characterization of biological specimens in their near-native state.

## 1. Introduction

Delivery of a specific volume of liquid is a crucial laboratory routine, which requires the accurate extraction of target liquid followed by transfer and dispersion. Nanofluidic devices have been developed to deliver low-volume liquid, such as nanopores,[1, 2] nanochannels and nanopipettes.[3-7] In particular, nanopipette is inspired by normal tip-shaped pipette with a nanoscale opening, and becomes a versatile tool for not only controlled delivery, but also biochemical analysis and single-cell manipulation.[8-10] Although nanopipettes could be used for liquid delivery, the volume of extracted liquid is typically in the range of microliters to milliliters,[11, 12] and the extraction by the electrowetting technique requires a significant amount of target liquid as the bath solution.[13] Considering that solution containing biomolecules or drugs is typically expensive and tiny-volume, a novel and low-cost method to directly extract the liquid at the sub-microliter scale is still being explored.

For biomedical research, exploring the interplay of biomolecules at the cellular level relies on the precise extraction of the hydrated samples, as well as the associated characterization techniques.[11] With the development of liquid cell nanofabrication, transmission electron microscopy (TEM) imaging of the liquid "sandwiched" between two ultrathin membranes has been proved to be feasible, and applied to varied biological specimens including cells and proteins.[14-16] In addition to the sealed liquid cell, the cryo-TEM approach has also been applied to attain structural information in the frozen hydrated state.[17] Despite the advances of TEM in acquiring images of the nanostructure, atom probe tomography (APT) is considered to be a unique technique providing three-dimensional (3D) chemical maps of nanostructure at near-atomic resolution (~0.3 nm laterally and ~0.1 nm in depth).[18] However, the fabrication of the liquid cell for APT is a technical challenge since the APT specimen is required to be an ultra-sharp tip shape with an end radius less than ~75 nm.[18] As such, the method to prepare the tip-shaped liquid cell has gained research interests. Recently, graphene has been demonstrated to encapsulate a specific volume of target liquid for APT,[19-21] and the graphene encapsulation method holds great potential to immobilize the liquid with tunable volume.

In this study, we propose a graphene nanopipette (GNP) for the precise extraction of liquid by encapsulating target liquid on the prefabricated metal tip with a single graphene membrane (Figure 1a and 1b). Simulation based on finite element analysis (FEA) is carried out to provide insights into the dynamics of the encapsulation process, followed by a series of controlled experiments. The results demonstrate that the encapsulating speed is the key contributing factor for a successful encapsulation, and confirm that the volume of the encapsulated liquid is highly controllable at zeptoliter precision by tuning the encapsulating speed and the number of the encapsulation rounds (Figure 1c). Furthermore, GNP is a new type of tip-shaped liquid cell that enables the characterization of single biomolecules in their hydrated state. Using isotope ($^{13}$C)-labelled glucose as an example, the compositions and the quantitative distribution of single glucose molecules in the solution have been revealed by APT in the laser-pulsed mode. Notably, using a straightforward laboratory setup, the graphene encapsulation method can be completed in a short period (~5 min) at room temperature and ambient pressure.

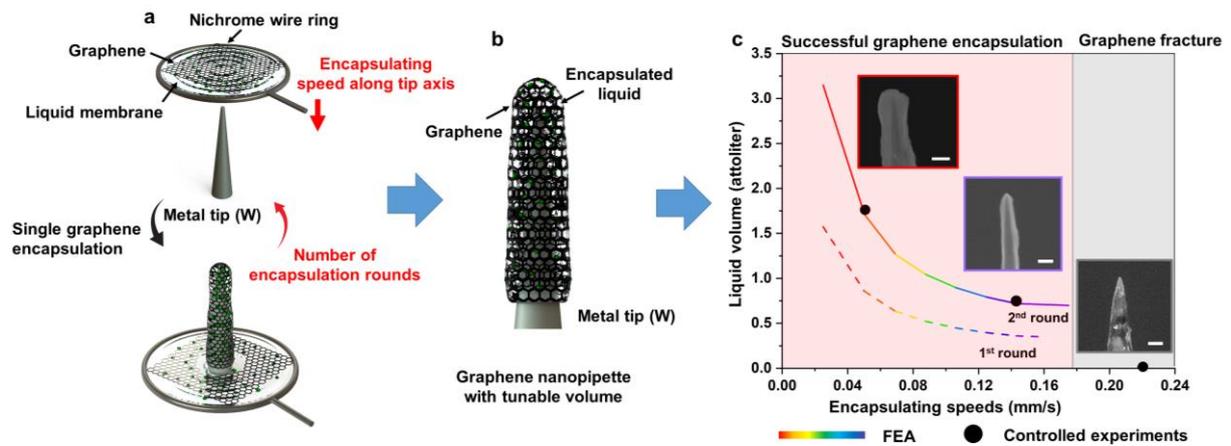

**Figure 1.** (a) Schematic of the fabrication of the graphene nanopipette (GNP) by encapsulating liquid on the prefabricated tungsten (W) tip with graphene. Inside the nichrome wire ring, the graphene membrane floats on the liquid membrane containing the sample solution (Figure S1). By lowering the wire ring through the prefabricated W tip using a specified speed along the tip axis, a small portion of graphene separates and encapsulates zeptoliters to attoliters of liquid on the W tip. Additional liquid volume is accumulated on the W tip after repeating the graphene encapsulation. (b) Schematic of the fabricated GNP with tunable volume. (c) Volume of the encapsulated sample (ferritin) solution precisely tuned by the encapsulating speed and the number of graphene encapsulation rounds, confirmed by FEA and the controlled experiments. Scanning electron microscopy (SEM) images of tested ferritin-containing GNP are shown in the insets, where scale bars represent 200 nm.

## 2. Results and discussions

2.1 Tuning the liquid volume at zeptoliter precision

Due to that observing the dynamic process of graphene encapsulation is extremely difficult, a finite element (FE) model (Figure S2), consisting of a graphene membrane, a liquid membrane and a tungsten (W) tip, is constructed to simulate the process of graphene encapsulation. It should be noted that the liquid membrane is simulated using the Coupled Eulerian-Lagrangian method.[22] Further details about the material settings, element types and boundary conditions have been provided in the Supporting Information. Outputs such as the maximum stress on the graphene membrane during encapsulation can be acquired. As shown in Figure 2a, the maximum stress on graphene occurs at the contact region when graphene starts to contact the W tip with the speed of 0.16 mm/s, and maintains an upward trajectory to 84 GPa. Subsequently, the graphene membrane starts to fold, and gradually cover the entire tip apex. The maximum stress (~74 GPa) is then relocated to the graphene boundary regions where the wrinkles are concentrated. Eventually, high stresses occurring at the boundary break the attached liquid membrane, and a limited amount of liquid is retained between graphene and the W tip. Due to the van der Waals force,[23] boundary regions of the graphene membrane can adhere to the tip surface, forming an excellent seal.

Based on the fracture strength of graphene (~100 GPa) reported in the literature,[24] the fracture pattern of graphene at a high encapsulating speed can be simulated. As such, the maximum speeds allowing graphene encapsulation without fracture could also be determined as a function of the radii of prefabricated W tips, providing a speed limit curve during practice (Figure 2b). Furthermore, the maximum allowed encapsulating speed has been found to increase linearly with the increase of the W tip radius. Controlled experiments have been performed to verify the determined speed limit curve, and the graphene membrane is shown to successfully encapsulate the sample (ferritin) solution on the W tip when the encapsulating speed is less than the maximum allowed value. For example, inset images in Figure 2b present both the simulation and experiment results of graphene encapsulation at different speeds, when the radii of W tips are constantly 25 nm. The graphene membrane has been fractured when the encapsulating speed (0.25 mm/s) is higher than the maximum allowed value (~0.17 mm/s), with no ferritin solution encapsulated. In contrast, the graphene membrane is shown to be intact at the encapsulating speed of 0.16 mm/s, forming a ferritin-containing GNP with predictable volume.

FEA further suggests that the volume of GNP can be tuned by the encapsulating speed and the number of graphene encapsulation rounds. As shown in Figure 2c, a 3D surface summarizes the results of FEA modeling the encapsulation of liquid on W tips (with constant radii of 25 nm) with varied encapsulation parameters. The volume of the encapsulated liquid has been found to increase in a nonlinear fashion with the decrease of encapsulating speed. Specifically, the volume dramatically increases when the encapsulating speed is below 0.08 mm/s. Also, the volume of the encapsulated liquid increases linearly with the increased number of graphene encapsulation rounds. The triangle and circle points in Figure 2c represent the results of controlled experiments, and scanning electron microscope (SEM) images of these tested GNPs containing sample (ferritin) solution are shown in Figure 2d. Each SEM image shows that a distinct layer of ferritin solution is observed on the W tip and the volume is at zeptoliter-scale as predicted. The experimental results fit well with the simulation results with 6% maximum deviation, confirming that the volume of the encapsulated liquid is highly controllable.

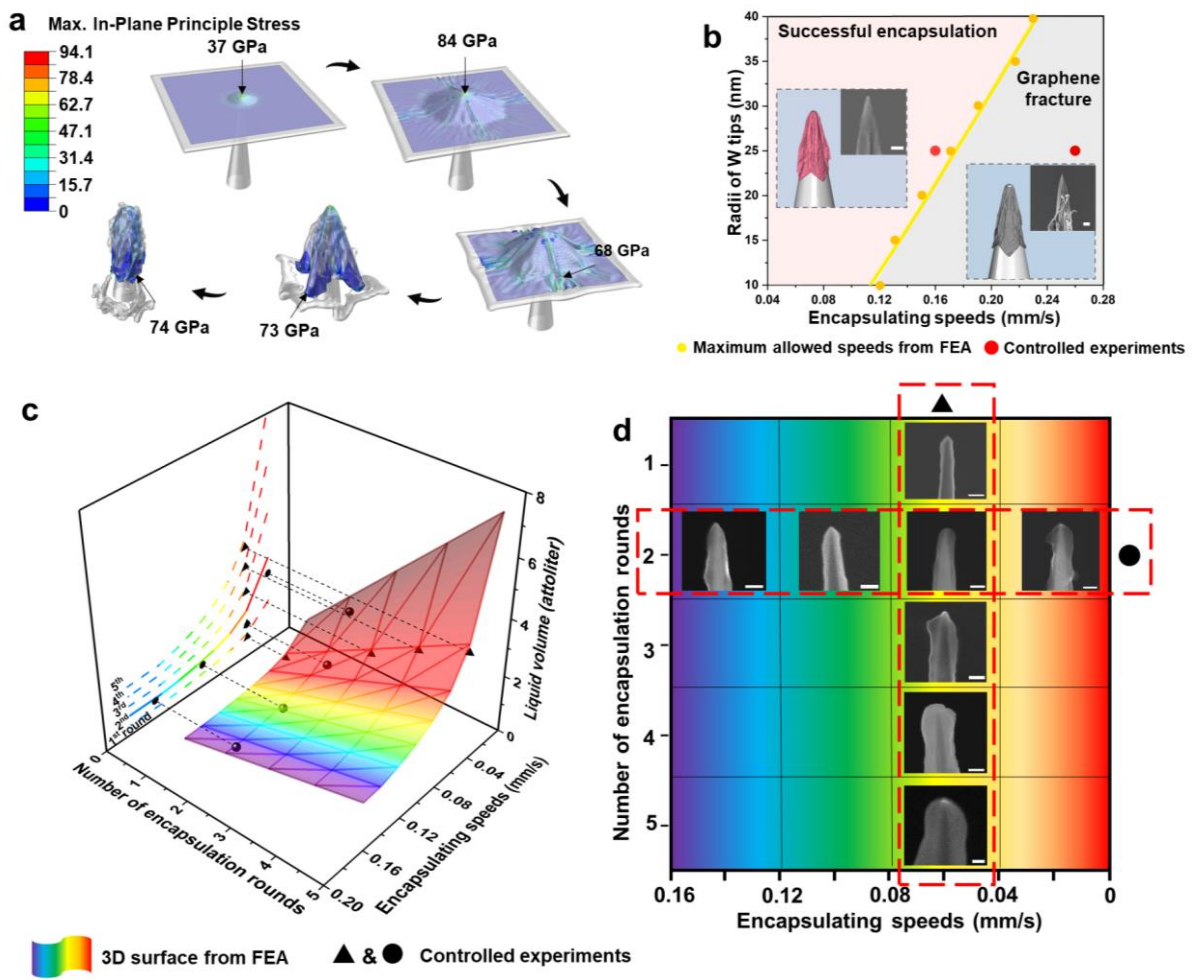

**Figure 2.** (a) Sequential images from simulation (FEA) showing the dynamics of graphene encapsulation. (b) Maximum allowed encapsulating speeds as a function of varied tip radii. (c) Three-dimensional (3D) surface formed by FEA revealing that the volume of GNP can be tuned by the encapsulating speed and the number of graphene encapsulation rounds. (d) SEM images

of the ferritin-containing GNP corresponding to the data points shown in (c). The scale bars are 200 nm across.

Due to the excellent sealing capability of the graphene membrane, releasing the encapsulated liquid is a technical challenge. Two methods have been attempted. Firstly, the end part of GNP, upon cryogenic freezing, can be milled using cryo-focused ion beam (FIB),[19] followed by immersing the frozen GNP in the deionized water at room temperature and ambient pressure. Secondly, the application of ultrasonication to the GNP immersed in the water could help release the encapsulated liquid,[25] while damages could occur in the released biomolecules. A reliable releasing method towards site-specific liquid delivery is still being investigated.

2.2 Graphene nanopipette enabled atom probe tomography of individual glucose molecules

Given that the volume of the encapsulated solution could be precisely controlled, the end radius of the GNP can be maintained less than 75 nm to allow the acquisition of 3D near-atomic chemical maps with APT.[18] Recently, single ferritins in solution have been successfully analyzed by APT after graphene encapsulation.[19, 20] In this study, we further extend this method for chemical mapping of single small molecules in solution. Using the $^{13}$C-labeled glucose molecules in solution as an example, the specimen was prepared by using single graphene encapsulation of solution on the prefabricated W tip at the encapsulating speed of 0.05 mm/s. After confirming that the geometry was suitable for APT imaging, the fabricated GNP was inserted into the atom probe instrument and cooled by a closed-cycle helium cryo-generator.[18] APT imaging was performed in the laser-pulsed mode, where field-evaporated ions were identified from the mass-to-charge-state ratio spectrum, followed by reconstruction of 3D chemical maps of the evaporated volume. According to the mass spectrum (Figure 3a and 3b), major peaks in addition to W and W oxide ions have been assigned to ions containing C, O and H (Table S1), which is consistent with the known composition of glucose.[26]

The field-evaporated ions have been reconstructed into the 3D chemical map using the tip-profile method, which defines the radius of the reconstructed chemical map as a function of the depth (Figure S3a) based on the SEM image of APT specimen (Figure S3b). As shown in Figure 3c, the chemical map consists of two distinct volumes, referring to the graphene-encapsulated glucose solution at the top and the base W tip. The W ions are the most abundant of detected ions, and $Ga^{2+}$ (34.5 Da) ions have been identified inside the reconstructed tip volume since the W tip was prefabricated by FIB with gallium ions. Notably, these ions are only detected 18 nm below the top of the evaporated volume, suggesting that the ions at the top originate from the graphene-encapsulated glucose solution. The transition from the encapsulated solution into the W substrate could also be revealed in the voltage and mass history (Figure S3c). Inside the reconstructed solution volume, numerical $H_2O$-related ions, including $OH^+$ (17 Da), $H_2O^+$ (18 Da) and $H_3O^+$ (19 Da), have been detected, confirming that the specimen maintains the frozen hydrated state. $Na^+$ (23 Da) ions are considered to originate from the supplements in the glucose solution. Many C-containing ions have been detected throughout the solution volume, including $C^{1+,2+}$, $C_{1-4}H_{0-3}^+$ and $C_{1-3}OH_{0-3}^+$. Given that unique peaks at 6.5 Da and 13 Da have not been previously observed in the mass spectrum of biological specimens acquired by laser-pulsed APT,[19, 27, 28] and the ratio of the count is approximately 1, the peaks at 6.5 Da and 13 Da can be identified as $^{13}C^{1+,2+}$, confirming that $^{13}C$-labelled glucose molecules have been evaporated during field evaporation. Furthermore, the corresponding chemical map shows that $^{13}C^{1+,2+}$ ions are only located in solution volume, demonstrating that APT imaging of GNP could be a novel experimental protocol for tracking isotopic labeling through single molecules at near-atomic resolution.

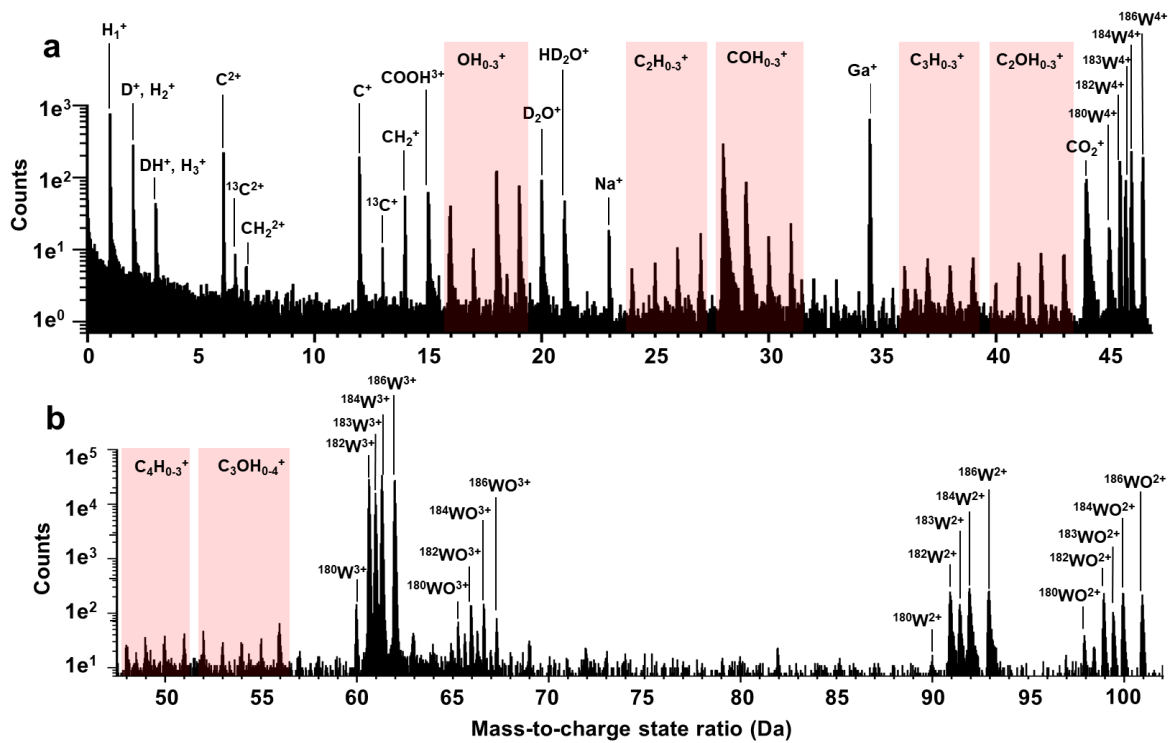

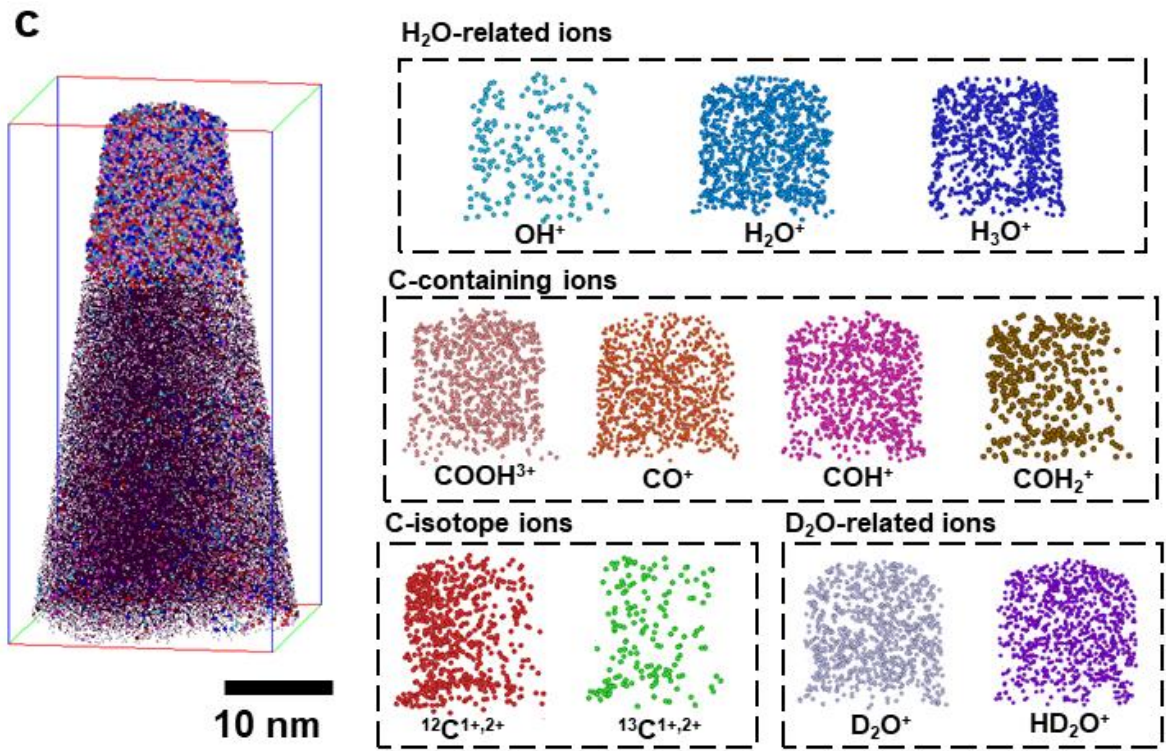

**Figure 3.** Mass-to-charge-state ratio spectrum of graphene-encapsulated solution containing $C^{13}$-labelled glucose molecules and $D_2O$ on the prefabricated W tip in the range of (a) 0-47 Da and (b) 47-101 Da. (c) 3D reconstruction of the APT specimen, revealing individual maps for different ionic species.

Considering that $H_2O$-related ions could originate from glucose, heavy water ($D_2O$) was pre-added into the liquid membrane and encapsulated together with $^{13}C$-labelled glucose molecules on the prefabricated W tip. $D_2O$-related ions including $D_2O^+$ (20 Da) and $HD_2O^+$ (21 Da) have been identified (Figure 3a), further confirming that the glucose molecules are in the hydrated state. Figure 4a is a proximity histogram according to the isoconcentration surface drawn at 48.4 at.% $^{180-186}W^{2-4+}$, revealing that $D_2O^+$ and $H_2O^+$ ions are distributed throughout the solution volume. In addition, the proximity histogram also reveals a W oxide film on the surface of the W tip. The oxide film can form when the W tip is exposed to the ambient environment after FIB milling.[29]

In order to locate single $^{13}C$-labelled glucose molecules, clustering analysis is carried out using the maximum separation algorithm in which ions are clustered if their spacing is smaller than a defined distance threshold ($d_{max}$).[30] The $d_{max}$ parameter has been determined to be 0.27 nm according to the first nearest neighbor (1NN) distribution of $^{13}C^{1+,2+}$ and C-containing ions

(Figure 4b), which is reasonable considering the theoretical C-C bond length (0.154 nm) and the detection efficiency of atom probe instrument (54%).[31] Based on the chemical structure ($^{13}CC_5H_{12}O_6$) and the determined $d_{max}$, each $^{13}C^{1+,2+}$ clustered with one or more C-containing ions within the distance threshold of 0.27 nm is considered to be the fragment of a single $^{13}C$-labelled glucose molecule. Figure 4c shows the located 114 clusters, representing possible sites of glucose. For example, the highlighted cluster includes one $^{13}C^+$, two $CO^+$ and two $COH^+$.

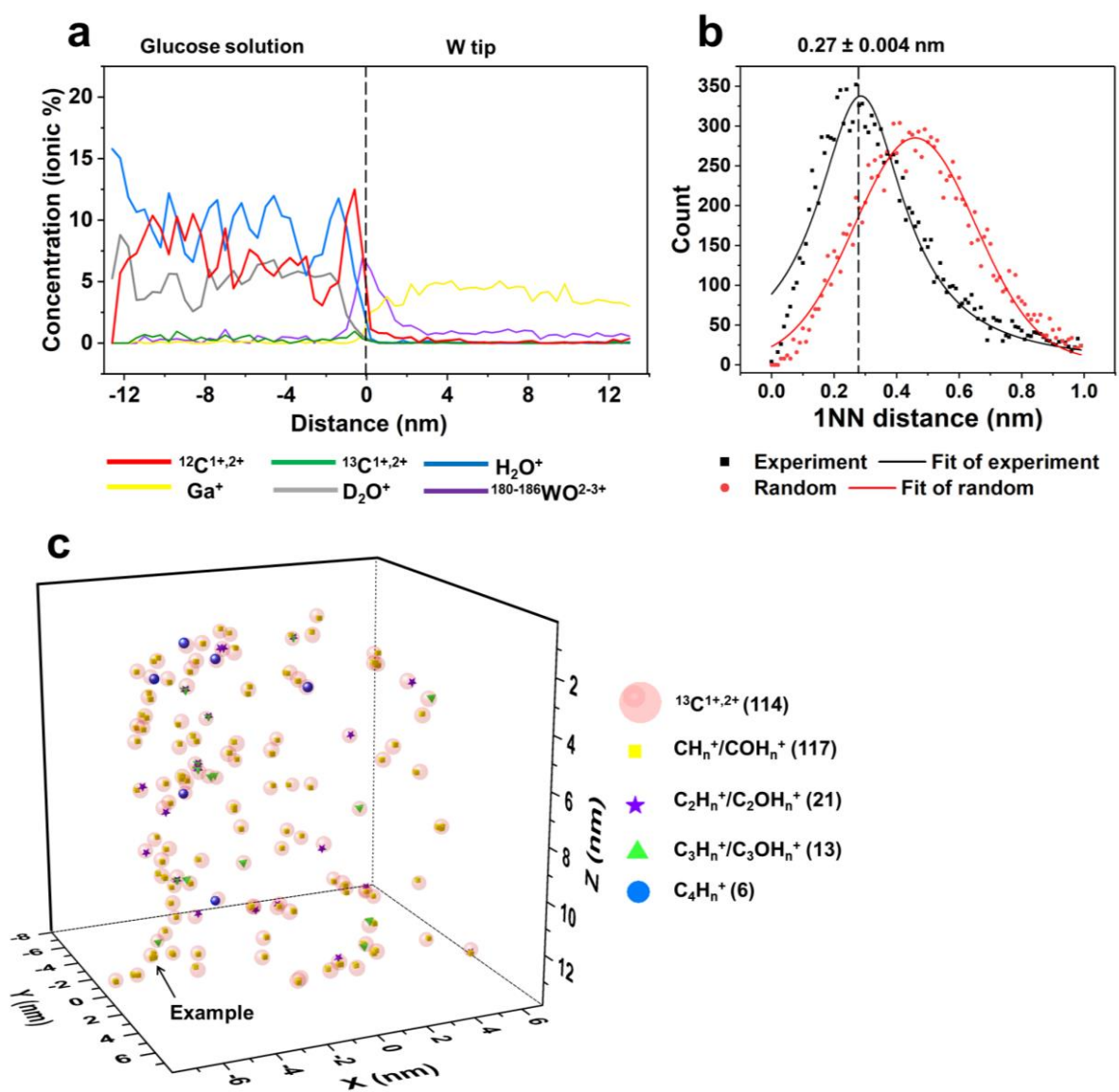

Figure 4. (a) Proximity histogram across the interface between the glucose solution and W tip, according to the isoconcentration surface of $^{180-186}W^{2-4+}$ (48.4 at.%). (b) First nearest neighbour (1NN) distribution of $^{13}C^{1+,2+}$ and C-containing ions (including $C^{1+,2+}$, $C_{1-4}H_{0-3}^+$ and $C_{1-3}OH_{0-3}^+$). The random distribution of corresponding ions is simulated by the 3D reconstruction

software (IVAS 3.8). (c) Identified $^{13}C$-labelled glucose molecules using clustering analysis, where the counts of each ionic species have been included in brackets.

## 3. Conclusion

The proposed GNP enables the extraction of liquid at zeptoliter precision, laying the foundation for zeptoliter-volume liquid delivery and near-native characterization. The GNP is prepared by encapsulating biomolecule solution on prefabricated metal tips with single graphene membranes. FEA has been performed to determine the maximum allowed encapsulating speeds according to the geometry of metal tips, and estimate the volume of the encapsulated liquid under specific encapsulation parameters, which has been verified by the controlled experiments. The volume of the GNP has been found to decrease as the encapsulating speed goes up, but increase with the increased number of encapsulation rounds, demonstrating the graphene encapsulation of liquid at zeptoliter precision. Furthermore, the GNP is considered to be a new type of tip-shaped liquid cell. Since the end radius of GNP can be tuned to be less than 75 nm, the GNP containing $^{13}C$-labelled glucose molecules dissolved into $D_2O$ has been imaged with laser-pulsed APT. A large number of $H_2O$-related and $D_2O$-related ions have been identified, confirming the molecules in the frozen hydrated state. The compositions of single $^{13}C$-labelled glucose molecules have been determined, as well as their quantitative distribution in the solution. In the future, the GNP can be introduced to deliver sub-attoliter-volume drug solutions, and correlate multiple characterization techniques, such as cryo-TEM and APT, to attain both structural and chemical information of biological specimens.

**Acknowledgments**

This study was partly funded by the Australian Research Council (DP180103955). C.Z. is supported by the start-up funding at Westlake University. This work was performed in part at the Melbourne Centre for Nanofabrication (MCN), Victorian Node of the Australian National Fabrication Facility (ANFF). The authors also acknowledge the use of instruments and science and technical assistance at the Monash Centre for Electron Microscopy and Monash Ramaciotti Cryo-EM platform (the Victorian Node of Microscopy Australia), and Nanjing University of Science and Technology's Chinese Centre of Excellence for Atom Probe Tomography.